\RequirePackage{fix-cm}
\pdfoutput=1 
\documentclass{svjour3}                    % onecolumn (standard format)
\smartqed  % flush right qed marks, e.g. at end of proof

\def\makeheadbox{\relax}
\usepackage[utf8]{inputenc}
\usepackage[english]{babel}

\usepackage{geometry}
\usepackage{marginnote}

\usepackage{graphicx}
\usepackage[numbers,sort&compress]{natbib}
\usepackage{placeins}
\usepackage[caption=false]{subfig}

\newcommand{\figref}[1]{\figurename~\ref{#1}}
\newcommand{\tabref}[1]{\tablename~\ref{#1}}

\journalname{International Journal of Computer Assisted Radiology and Surgery}

\usepackage[pdfborder={0 0 0}]{hyperref} % keep this package last

\usepackage{xcolor}

\begin{document}

\title{Manual segmentation versus semi-automated segmentation for quantifying vestibular schwannoma volume on MRI%\thanks{Grants or other notes
%about the article that should go on the front page should be
%placed here. General acknowledgments should be placed at the end of the article.}
}
%\subtitle{Do you have a subtitle?\\ If so, write it here}

%\titlerunning{Manual versus semi-automated segmentation}        % if too long for running head
\titlerunning{Manual versus semi-automated vestibular schwannoma segmentation}

\author{Hari McGrath       \and
        Peichao Li          \and
        Reuben Dorent       \and
        Robert Bradford    \and
        Shakeel Saeed      \and
        Sotirios Bisdas    \and
        Sebastien Ourselin  \and
        Jonathan Shapey     \and
        Tom Vercauteren}

%\author{Hari McGrath \and Peichao Li \and Reuben Dorent \and Robert Bradford \and Shakeel Saeed \and Sotirios Bisdas \and Sebastien Ourselin \and Jonathan Shapey \and Tom Vercauteren}

%\authorrunning{Hari McGrath et al.} % if too long for running head

\institute{H. McGrath
              \at GKT School of Medical Education, King's College London, London, UK\\
              \email{hari.mcgrath@kcl.ac.uk}
           \and
           H. McGrath \and P. Li \and R. Dorent \and S. Ourselin \and J. Shapey \and T. Vercauteren
              \at School of Biomedical Engineering \& Imaging Sciences, King’s College London, London, UK
           \and
           R. Bradford
              \at Queen Square Radiosurgery Centre (Gamma Knife), National Hospital for Neurology and Neurosurgery, London, UK
           \and
           R. Bradford \and S. Saeed \and J. Shapey
              \at Department of Neurosurgery, National Hospital for Neurology and Neurosurgery, London, UK 
           \and
           S. Saeed
              \at The Ear Institute, UCL;
              %, London, UK
              %\at
              The Royal National Throat, Nose and Ear Hospital, London, UK
           \and
           S. Bisdas
              \at Neuroradiology Department, National Hospital for Neurology and Neurosurgery, London, UK
           \and
           J. Shapey
              \at Wellcome/EPSRC Centre for Interventional and Surgical Sciences, UCL, London, UK
           }

%\date{Received: date / Accepted: date}
\date{}
% The correct dates will be entered by the editor

\maketitle

\begin{abstract}~\\
\textbf{Purpose}
Management of vestibular schwannoma (VS) is based on tumour size as observed on T1 MRI scans with contrast agent injection. Current clinical practice is to measure the diameter of the tumour in its largest dimension. It has been shown that volumetric measurement is more accurate and more reliable as a measure of VS size. The reference approach to achieve such volumetry is to manually segment the tumour, which is a time intensive task. We suggest that semi-automated segmentation may be a clinically applicable solution to this problem and that it could replace linear measurements as the clinical standard.\\
\textbf{Methods}
Using high-quality software available for academic purposes, we ran a comparative study of manual versus semi-automated segmentation of VS on MRI with 5 clinicians and scientists. We gathered both quantitative and qualitative data to compare the two approaches; including segmentation time, segmentation effort and segmentation accuracy.\\ 
\textbf{Results}
We found that the selected semi-automated segmentation approach is significantly faster (167s versus 479s, $p<0.001$), less temporally and physically demanding and has approximately equal performance when compared with manual segmentation, with some improvements in accuracy. There were some limitations, including algorithmic unpredictability and error, which produced more frustration and increased mental effort in comparison to manual segmentation.\\
\textbf{Conclusion}
We suggest that semi-automated segmentation could be applied clinically for volumetric measurement of VS on MRI. In future, the generic software could be refined for use specifically for VS segmentation, thereby improving accuracy. 

% \PACS{PACS code1 \and PACS code2 \and more}
% \subclass{MSC code1 \and MSC code2 \and more}
\end{abstract}

\section{Introduction}
\label{intro}
Vestibular schwannoma (VS) is a benign tumour of the vestibulocochlear nerve arising within the cerebellopontine angle, deep inside the cranium. It accounts for approximately 6-8\% of all intracranial neoplasms and has a prevalence of around 0.02\% of the population 
\citep{lin2005prevalence}. 
Patients may present with a variety of symptoms including hearing loss, balance problems, vertigo, dizziness and headache among others \citep{pinna2012vestibular}. 
Diagnosis is usually made on a Magnetic Resonance Imaging (MRI) scan with intravenous contrast demonstrating a homogeneously-enhancing lesion within the internal acoustic canal that may also extend into the intracranial cavity~\citep{nguyen2019vestibular}. Grading of tumours is performed according to radiographic characteristics indicating tumour extent and size and is used to guide treatment \citep{koos1998neurotopographic}. Patients with small or asymptomatic tumours are usually managed conservatively with serial surveillance scans. Small or medium-sized tumours deemed suitable for treatment can be treated effectively and safely with stereotactic radiosurgery (SRS) \citep{lunsford2005radiosurgery} but larger tumours are usually managed with surgery.

Measuring the size of a VS on MRI is important in guiding treatment or monitoring growth patterns. There are several methods for measuring tumour size but the most common technique is to measure diameter at the tumour's widest point 
\citep{yoshimoto2005systematic, kanzaki2003new, shapey2018standardised}. However, this approach is prone to measurement inaccuracies. 
Volumetric measurement is a solution to this problem \citep{van2009follow}. Volumetric analysis offers a more accurate representation of the tumour \citep{lees2018natural} and could significantly aid the management of these patients. Segmentation (contouring) is already used in the planning of gamma knife SRS treatment. Segmentation also provides a means of performing volumetric measurement of the tumour. Compared with two-dimensional measurements, it may be used more accurately for the active surveillance of VS. Volumetric measurement has been used to predict recurrence in patients with residual tumours following surgical intervention \citep{vakilian2012volume}, to measure change in tumour size following SRS treatment \citep{yu2000sequential} and to predict hearing preservation following SRS treatment \citep{gjuric2007hearing}. There are three main methods of volumetric analysis: manual segmentation, semi-automated segmentation and automated segmentation. Manual segmentation involves comprehensively labelling the 3D structure in each 2D slice. It is a time-intensive task with relatively low inter- and intra-individual reliability and has not been widely employed in clinical practice. 

Automated segmentation has been applied successfully to MR imaging for a wide range of brain tumours \citep{zou2004three}. Automated segmentation may be accurate in the assessment of tumour progression and in overall survival prediction in glioma \citep{bakas2018identifying, menze2015brats} as well as for the clinical assessment of biomarkers in glioma \citep{booth2020ml}. For VS imaging, automated segmentation has been applied with positive results \citep{wang2019deep, shapey2019artificial} and there is growing interest in the field \citep{george2020auto}. An automated segmentation tool could also improve clinical workflow and operational efficiency during the planning of stereotactic radiosurgery (SRS) by using the tool as an initialisation step in the process. However, automated approaches are, for the most part, not fully validated and are confined to academic use. Furthermore, some tumours display heterogeneous enhancement including the $4\%$ of VS tumours that may be cystic, which can lead to inaccurate segmentation when automated methods are applied \citep{mazzara2004brain}.

Semi-automated segmentation has been shown to be a more reliable option for the analysis of VS on MRI scans 
\citep{mackeith2018comparison}. However, there has been no previous analysis of cognitive load or user experience of VS segmentation. When using semi-automated methods, segmentation time and repeatability may be improved when compared with manual segmentation \citep{wang2019deepigeos,wang2019interact,birkbeck2009interactive,chae2017semi}. Compared with fully automatic segmentation, results may be more accurate \citep{bakas2018identifying} and are more acceptable to clinicians due to increased transparency in the segmentation process \citep{gordillo2013state}. Currently proposed methods require user input for one or more of the following steps: segmentation parameters, feedback or evaluation, including refinement and validation of the segmentation. There is little material in the literature regarding user experience of interactive segmentation in brain imaging, despite the intention to pursue clinical translation in the field \citep{khademi2019whole,shaver2019optimizing}.

A number of software packages are academically available for medical image segmentation spanning a variety of different methods. For manual segmentation, ITK-SNAP\footnote{\url{http://www.itksnap.org}}~\citep{yushkevich2006user} is a widely-used open-source software library with manual, semi-automated and automated segmentation offerings. 3D slicer\footnote{\url{http://www.slicer.org}} has the standard offerings of image viewing and analysis tools, along with a variety of downloadable packages for semi-automated and automated segmentation
\citep{fedorov2018seg}. MRIcron\footnote{\url{https://people.cas.sc.edu/rorden/mricron/}} is a package of image viewing and manual segmentation tools. For semi-automated segmentation, ImFusion Labels (ImFusion, Munich, Germany) is a recent commercial-grade package with academic licensing options.

We present the findings of a proof of concept study using combined quantitative and qualitative analysis, comparing manual segmentation with semi-automated segmentation of VS on MRI. We hypothesise that semi-automated segmentation is faster than manual segmentation with a comparable performance. In this study we also compare the user experience of two software suites, including that of clinicians and senior researchers.

\section{Materials and Method}
\label{sec:materials_and_methods}
We selected four tumours from our database for the study (See \tabref{tab:tumour_characteristics}). All four patients had previously undergone Gamma Knife SRS treatment
\citep{boari2014gamma}. The images were representative of a variety of tumour sizes and shapes encountered in clinical practice. We selected two small and two moderate-sized tumours (See \tabref{tab:tumour_characteristics}).
The ground truth measurements were made prior to the study by the treating skull base neurosurgeon and stereotactic radiosurgery physicist using Gamma Knife planning software (Leksell GammaPlan, Elekta, Sweden). The images used in this study were all contrast-enhanced T1-weighted scans with $0.4mm\times0.4mm$ in-plane resolution, in-plane matrix of $512\times512$ and $1.5mm$ slice thickness. All cases included an extracanalicular (intracranial) component and none of the tumours had a cystic component. 
Patients with multiple tumours were excluded. 

\begin{table}
% table caption is above the table
\caption{Tumour characteristics according to commonly used criteria for representing tumour size and extent}
\label{tab:tumour_characteristics}       % Give a unique label
% For LaTeX tables use
\begin{tabular}{llll}
\hline\noalign{\smallskip}
Tumour Identifier & Volume ($mm^3$) & Largest Diameter ($mm$)  
\\
\noalign{\smallskip}\hline\noalign{\smallskip}
VS\_1 & 623 & 15.1 
\\
VS\_2 & 1050 & 20.5 
\\
VS\_3 & 3590 & 25 
\\
VS\_4 & 975 & 17 
\\
\noalign{\smallskip}\hline
\end{tabular}
\end{table}

We selected ITK-SNAP for manual segmentation since this offered the most intuitive user interface. In our group it was also the most widely used library for manual segmentation. We selected ImFusion Labels for semi-automated segmentation since this was a recent software with a good selection of machine learning tools and a high-quality user interface. It was made available to our group through an academic license. 

Five observers, including two medical students, two biomedical engineers and one neurosurgeon, performed manual and semi-automated segmentation on each of the four scans. The participants had a variety of experience with segmentation. Three participants were inexperienced segmenters (with no or limited previous experience) and two were experts in medical image segmentation, with multiple years experience of medical image segmentation. Three had previous experience using ITK-SNAP, one of whom had limited experience of using ImFusion Labels.  

\subsection{Study Design}
%\label{subsec:study_design}
A training period was included for each study participant at the start of the study and for each software library, using a training data set which was not part of the study. This training period was standardised to 10 minutes for each participant and included an initial demonstration from the study lead followed by a trial run for each participant. During the training period, participants were free to ask questions relating to the segmentation. The trial runs were not included in the results or the analysis. Participants were advised on the optimal tools to use in each software library. This training period was adapted based on the needs and previous experience of the participant, such that no demonstration was given for those participants well-versed in the use of the software library.

\begin{figure}[htbp!]
  \includegraphics[width=\textwidth]{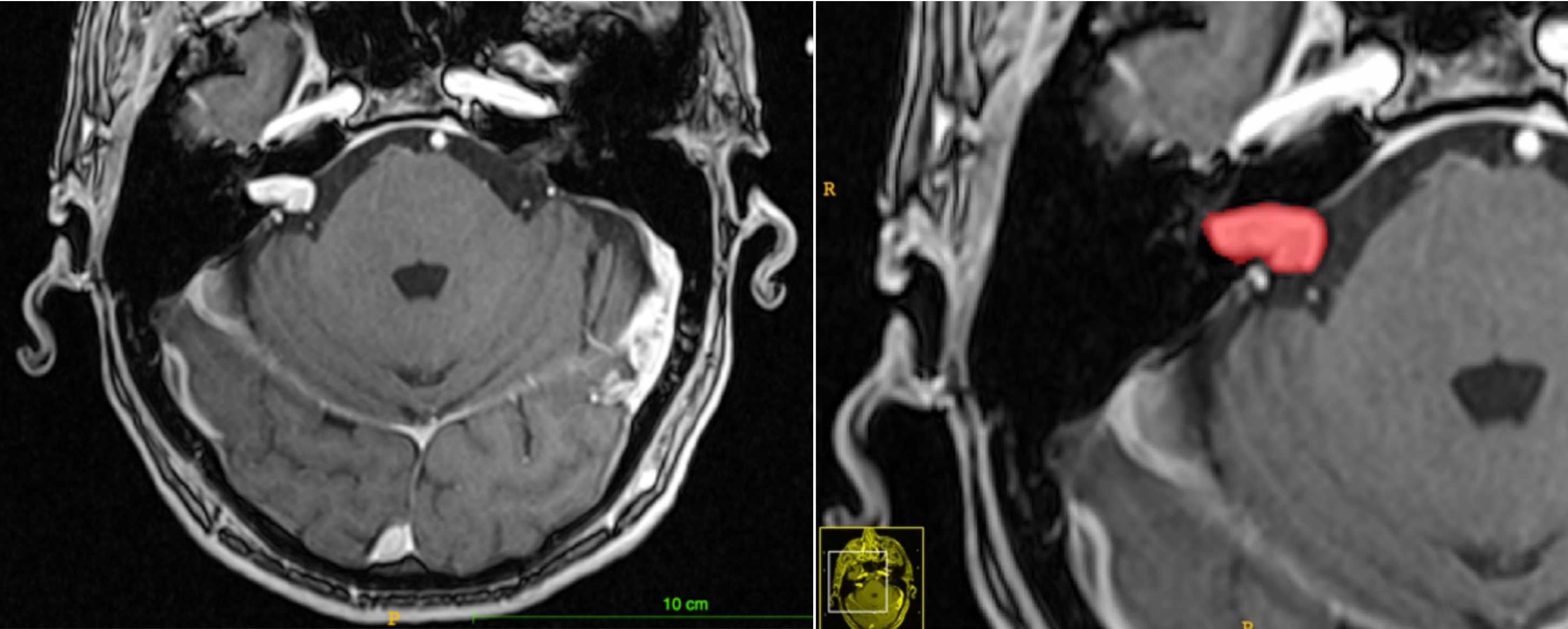}
\caption{Example of a tumour, pre- and post-segmentation, represented in ITK-SNAP. This was tumour 'VS\_1', classed as a small tumour with limited extracannalicular extension}
\label{fig:vs-pre-post-surg}       % Give a unique label
\end{figure}

In ITK-SNAP, participants used the polygon drawing tool to outline tumour boundaries in each slice and fill in the tumour volume (See \figref{fig:vs-pre-post-surg}). The paintbrush tool was used to make small alterations as needed. A time limit of ten minutes per segmentation was provided in order to standardise the process according to arbitrary mock-clinical parameters.
 
\begin{figure}[htbp!]
\sidecaption
  \includegraphics[width=0.7\textwidth]{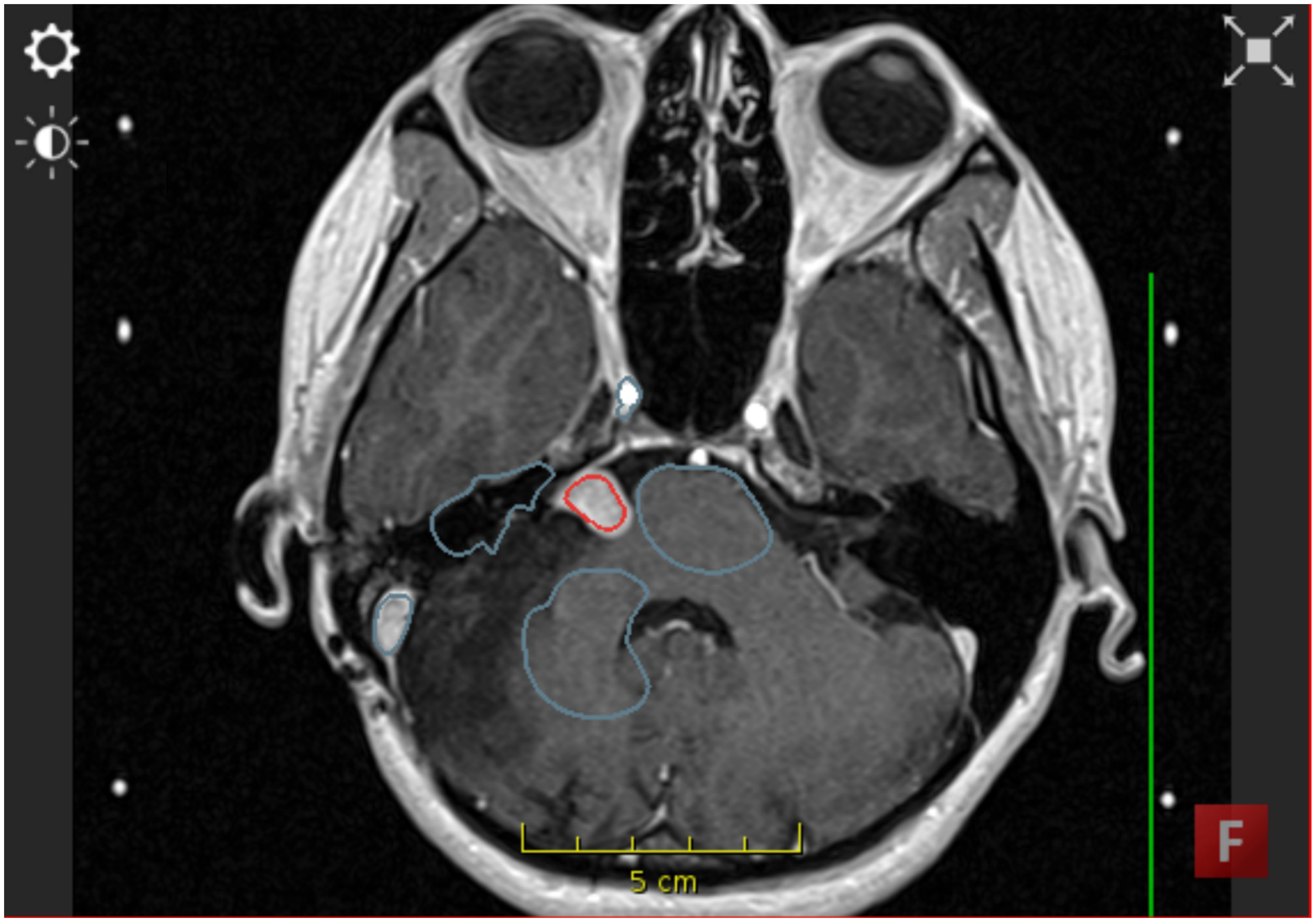}
\caption{Segmentation in ImFusion Labels using background labels (blue) and foreground labels (red) to demarcate tumour and non-tumour tissue}
\label{fig:seg-imfusion}       % Give a unique label
\end{figure}

In ImFusion Labels, participants used the ‘Interactive Segmentation’ module (See \figref{fig:seg-imfusion}). They were advised to first draw background labels which included structures of a variety of intensities (e.g. bone, dura, healthy brain). After the first iteration of the segmentation, participants were advised to only undertake two alterations in the segmentation. This was determined to produce optimum results while creating an incentive to complete the task in a time-pressured manner.

A document containing participant instructions is included as Online Resource~1. A video depicting segmentation in ITK-SNAP is included as Online Resource~2. A video depicting segmentation in ImFusion Labels is provided as Online Resource~3. 

\subsection{Qualitative Data Collection}
%\label{subsec:qualitative_data_collection}
The NASA Task Load Index (TLX) \citep{Hart1988DevelopmentON} questionnaire was performed at the end of the study to quantify user effort for each method of segmentation. The TLX scores different aspects of a task on a graded scale from 1-21, including effort, frustration and performance. It can be found as 
\tabref{tab:TLX_questionnaire} in the appendix. The TLX was used as a relative comparator of the libraries, rather than as an absolute scale. For data analysis we processed the raw TLX data. This may be a more reliable use of the TLX compared with using part two to calculate an overall weight-adjusted score~\citep{bustamante2008measurement}. 

We performed short post-segmentation interviews to explore the participants’ experiences of the different toolboxes. The questions were based around themes, which included ‘segmentation experience’, ‘toolbox’ and ‘study design’.
\tabref{tab:questionnaire} in the appendix details the questions asked of each participant. 
Participants were asked about each software library separately. Data was collected in shorthand form by the study lead during the interview and then expanded following the interview.

\subsection{Quantitative Data Collection and Analysis}
%\label{subsec:quantitative_data_collection}
The time taken to perform the segmentation was measured from the time of launching the software to the time of closing the software following the segmentation.
A paired t-test was performed on this data to calculate the p-value as well as the confidence intervals.
We quantified segmentation accuracy by comparing the segmentations in each software with the ground truth data in order to establish a comparative analysis.
We calculated the Dice Coefficient (Dice) since this is a standard comparative measure of radiological data \citep{menze2015brats, milletari2016vnet}. We also calculated relative volume error (RVE) and average symmetric surface distance (ASSD) for each segmentation. We performed subgroup analysis on both the time and accuracy data. We took the two more experienced segmenters and compared results from these individuals against the three less experienced segmenters.

\section{Results}
\label{sec:results}
Segmentation time was significantly faster in ImFusion Labels. In terms of TLX data, ITK-SNAP was more time demanding and physically demanding whereas ImFusion was more mentally demanding and frustrating. The performance, in terms of accuracy, and overall effort of the libraries was comparable. Qualitatively, participants preferred the control that ITK-SNAP offered, however some did not like the time demand. ImFusion was a good tool for rapidly estimating tumour volume, but there were frustrating errors produced in complex tumour segmentation.

\begin{figure}[hbp!]
\subfloat[]{\label{fig:seg-time-comp}\includegraphics[width=0.49\textwidth]{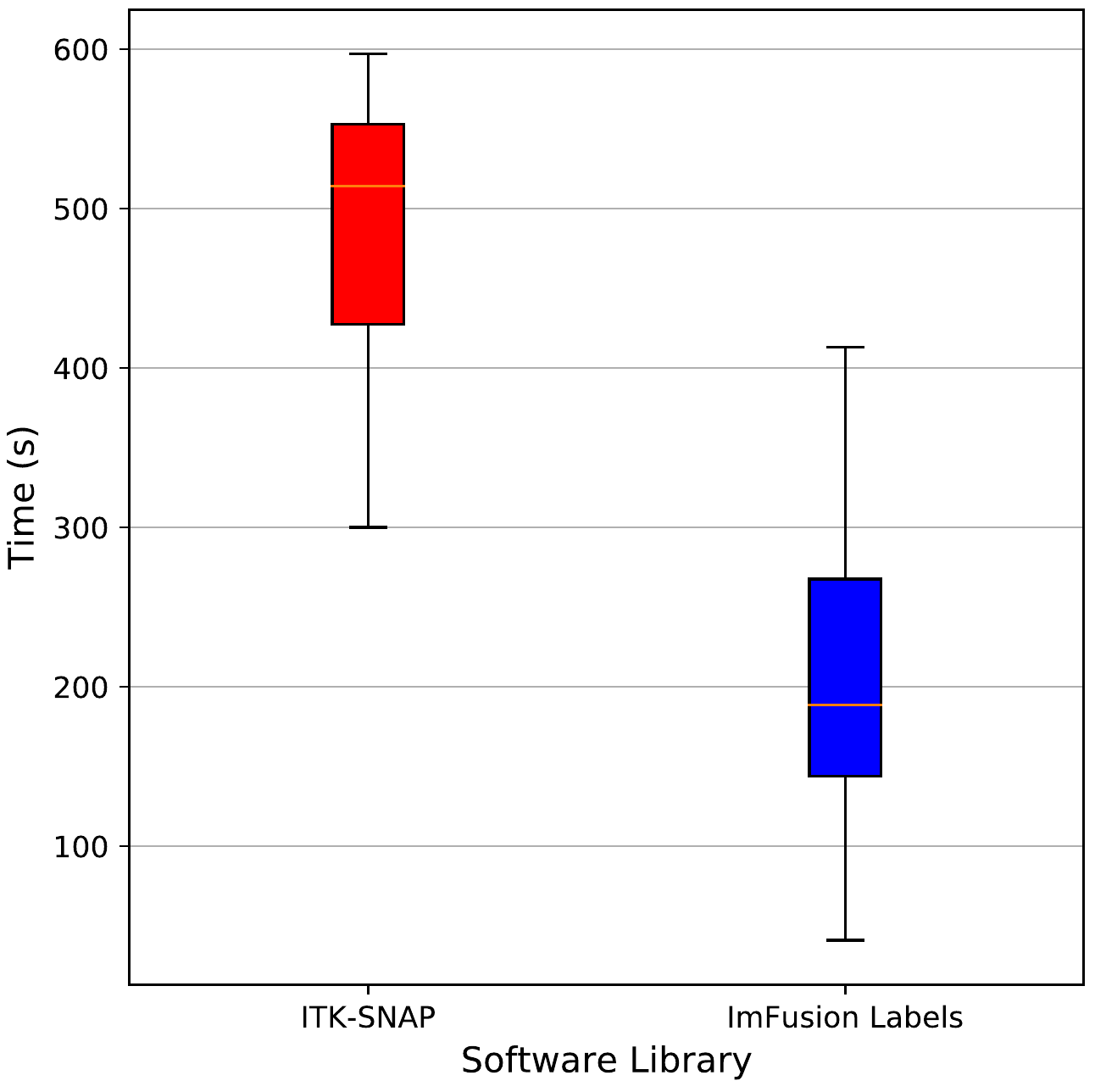}}
~
\subfloat[]{\label{fig:seg-dice-comp}\includegraphics[width=0.46\textwidth]{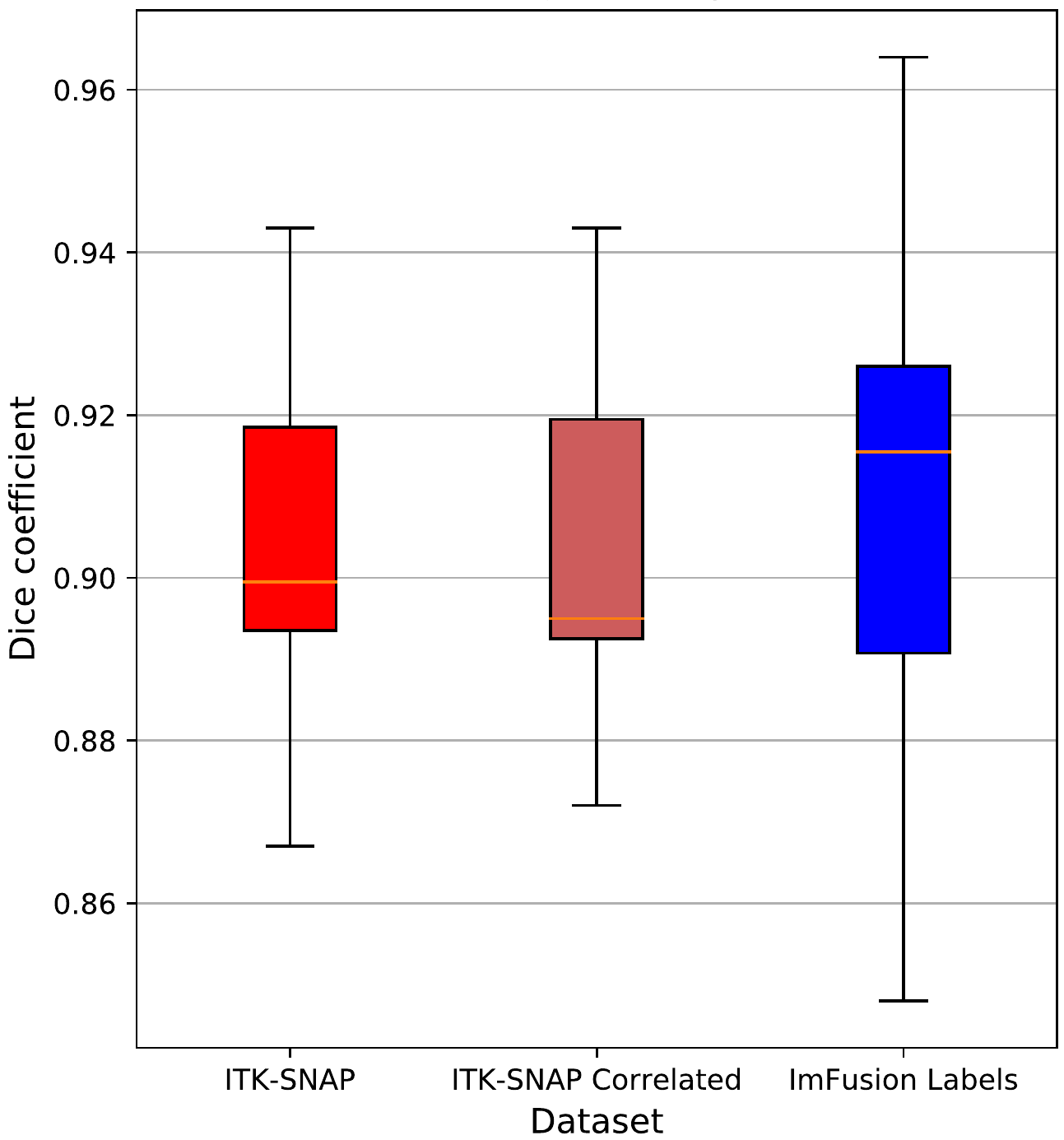}}
\caption{(a) Comparison of segmentation time between the two software libraries; (b) Spread of Dice scores in ITK-SNAP as compared to ImFusion Labels. The ``ITK-SNAP Correlated" plot only takes into account the data which corresponds to the one from ImFusion labels that we still had access to (after data loss had occurred).}
\end{figure}

\subsection{Time}%
Between the two libraries, segmentation in ImFusion Labels was significantly faster than ITK-SNAP. 
The mean segmentation time (ST) in ITK-SNAP was 479s (95\% CI 439 – 519) while the mean ST in ImFusion Labels was 168s (95\% CI 168 – 249), with a p value of $<0.001$ (See \figref{fig:seg-time-comp}). 
There was no observed difference in segmentation time between the less experienced individuals and the more experienced individuals.

\subsection{Accuracy}

The user-generated segmentation dataset was compromised during the study, resulting in half of the ImFusion data being unavailable for analysis of segmentation accuracy.
On the remaining data, we observed comparable accuracy between the two libraries, with a Dice score range of 0.848-0.964 for ImFusion compared with a range of 0.867-0.943 for ITK-SNAP. Compared with segmentations in ITK-SNAP, segmentations in ImFusion Labels were more similar to the ground truth data in terms of Dice (0.913 vs 0.902, p=0.301), RVE (0.0723 vs 0.124, p=0.245) and ASSD (0.381 vs 0.419, p=0.349) as illustrated in \figref{fig:seg-dice-comp}.
In our subgroup analysis the two cohorts achieved similar levels of accuracy for manual segmentation in ITK-SNAP. The experienced cohort achieved more accurate Dice scores (0.901 vs 0.899, p=0.533), and RVD scores (0.155 vs 0.104, p=0.312) while the inexperienced cohort achieved more accurate ASSD scores (0.417 vs 0.420, p=0.936) when compared with ground truth data. However, none of these differences were statistically significant.                                                                                                  

\subsection{NASA TLX score}
The TLX scores showed a trend towards ITK-SNAP being the more physically and temporally demanding approach (+6 and +3.4-point scores on average respectively), while ImFusion tended to be more mentally demanding and worse in terms of perceived performance (-7.8 and -2.4 points on average respectively). All participants graded ImFusion as being more frustrating, with a +7.4-point greater score on average. All participants also graded ImFusion as being more mentally demanding, with a +7.8 greater score on average. ITK-SNAP was graded as being more physically demanding by all but one participant. Less experienced raters tended to score the segmentation performance of ImFusion higher than more experienced raters. Overall effort was slightly greater (+2.4 points on average) in ImFusion. 
\begin{figure}[htbp!]
  \includegraphics[width=\textwidth]{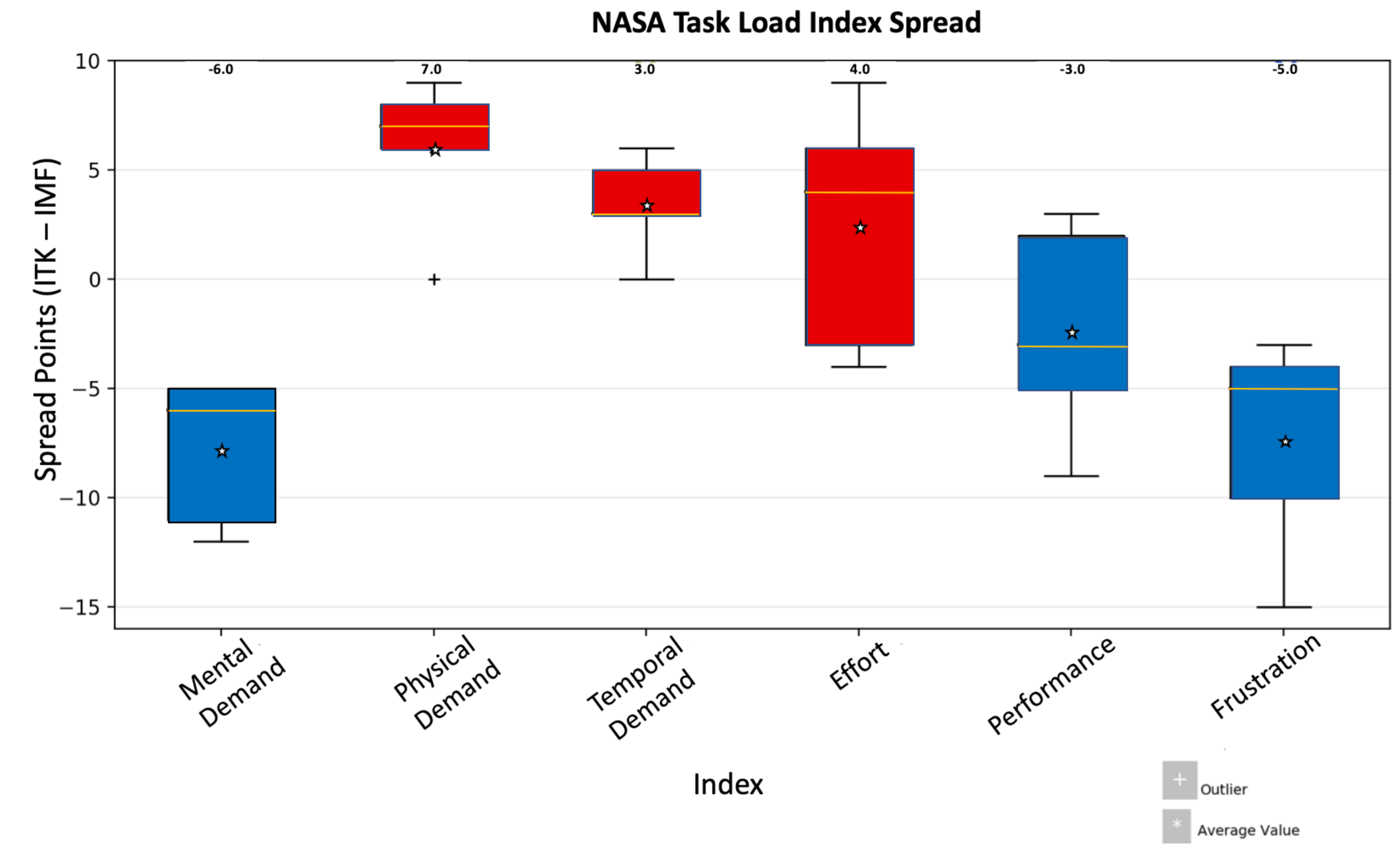}
  \caption{Relative NASA TLX spread data. The ImFusion score was subtracted from the ITK-SNAP score for each participant and combined for each index to show spread of data across participants. Positive values represent a greater score for ITK-SNAP, while negative values were greater for ImFusion. The scores at the top indicate the median value, while the colours represent the software which the mean value favoured. Blue indicates a mean which favoured ImFusion labels, while red indicates a mean value which favoured ITK-SNAP}
  \label{fig:5}
\end{figure}

\subsection{Interview Data}

ITK-SNAP was the preferred choice for highly accurate segmentation, whilst one participant recommended ImFusion as a ‘rough volumetric estimate’. All participants cited the improved performance of the ImFusion algorithm with ‘simple’ tumours i.e. those which were highly contrast-enhancing, homogeneous with well-defined boundaries and no or minimal adjacent high contrast structures, such as blood vessels or dura. However, for complex tumours the algorithm often made small, but frustrating, errors in segmentation - “[the algorithm] threw up errors which required a complete restart”. Occasionally non-tumour areas were included, and tumour areas were not included. There was generally no way to fix this using the tool. One participant complained that in these more challenging cases, the algorithm was “a one-trick pony… if you make alterations to the initial segmentation you may worsen it.” Participants commented on the ‘unpredictability’ of the algorithm and the lack of transparency as being a significant problem in solving these issues. In ITK-SNAP the majority of participants cited the need to compromise between thoroughness and timing of segmentation. One stated “I am a perfectionist… if we were not timed, [the segmentation] would take me much longer.”  In terms of study design, participants found the instructions clear and found it “helpful to have someone here to explain and provide feedback [during the training period]”. 
A full breakdown of the qualitative data taken from interviews is provided in the appendix (See \tabref{tab:questionaire-analysis}).

\section{Discussion}
\label{sec:discussion}
In this paper we sought to compare manual segmentation to semi-automated segmentation on several variables, both quantitative and qualitative, for segmentation of VS. It is widely published that semi-automated segmentation may reduce the time taken to perform segmentation \citep{fernquest2018segmentation,seuss2017development,ma2010repeatability}.
We showed that semi-automated segmentation is significantly faster and has comparable performance when compared with manual segmentation for volumetric analysis of VS. This would suggest good viability for this approach in clinical practice, where time constraints may restrict which methods are used. However, this study does have some limitations. 

In terms of performance, both semi-automated and manual segmentation were highly accurate when compared with ground truth data and there was no statistically significant difference between the two methods. In terms of clinical applicability, any differences between the two may also be clinically insignificant, thereby making semi-automated segmentation a desirable option.
The involvement of inexperienced segmenters may reduce the validity of the conclusions we can draw. However, we observed a high degree of similarity in accuracy data for the experienced segmenters when compared with the inexperienced segmenters, suggesting that there was no compromise on data quality due to the inclusion of less-experienced participants.

In interview, some participants suggested that the segmentation in ImFusion produced significant errors in complex tumours. The Dice scores, however, indicated a high degree of accuracy in these segmentations. One explanation for this inconsistency in perception versus result may be attributable to a finer margin for error applied to the analysis of segmentations in ImFusion. Participants spent, on average, 479s on each segmentation in ITK-SNAP, compared with 168s in ImFusion. This time discrepancy may have led to a higher acceptance threshold for the segmentation in ImFusion, and small mistakes may have been picked up more readily.   

In terms of effort measures, the NASA TLX was a useful tool. However, one limitation is that the system was used as a relative measure of effort between the different software libraries used for the study. Therefore, the absolute values offered by participants may not be an accurate measure of absolute effort and would therefore provide unreliable data for inter-rater comparison. We compared the inter-rater scores by subtracting the ImFusion scores from the ITK data for each participant. We would therefore suggest the use of the full TLX as opposed to the Raw TLX to overcome these issues.

We chose to state the segmentation goal as what would be clinically, or personally, acceptable to the participants. In this way, we felt that participants would apply the same requirements to both libraries. In some cases, the opposite was true. A very thorough approach was employed by some participants in ITK-SNAP, but in ImFusion Labels they used a crude approach. This difference in perceived goals may have introduced bias in the time and effort of segmentation. This challenge could be avoided in future by clearly stating the goals of the segmentation, whether targeting accuracy or speed. 

One constraint on semi-automated segmentation lies in usability of the tools. In this study, a common point of feedback was that the algorithm was inconsistent and unpredictable in its segmentation. Some users found this tedious and had to restart when the algorithm produced errors. In the literature, a commonly cited limitation in clinical application is algorithmic transparency 
\citep{kelly2019key}. Users did not understand what the algorithm did and why. ImFusion Labels is a generic library and has wide applicability in medical imaging. A solution to this issue may be to refine an algorithm specifically for VS segmentation.

There is very little qualitative data in the literature on the use of segmentation tools. Qualitative data are particularly important given the current interest in clinical translation of AI tools, which must be robust, easy to use and accurate~\citep{kelly2019key}. As far as we can see, this is the first paper to use a mixed quantitative and qualitative format to compare semi-automated segmentation with manual segmentation in medical imaging. The small sample size of this study, in terms of participants and scans segmented, may limit the validity of the conclusions we can draw. One further challenge was in data representation for qualitative analysis, since none of the research team had previous experience of handling interview data. It may be useful to recruit this expertise in future studies. 

In terms of applicability to the current clinical workflow, semi-automated segmentation may assist in monitoring VS growth, especially in those patients with small tumours being managed conservatively with serial imaging
\citep{hughes2011expectant,halliday2018update,shapey2018standardised}. It has been established that volumetric measurement is superior to single-dimension diametric measurements for quantifying growth \citep{mackeith2018comparison,varughese2012growth}. Manual segmentation is not feasible in routine clinical practice due to the time-demanding nature of the task. We showed that semi-automated segmentation is less time-demanding, less physically-demanding and of comparable performance. 

In the future, it is hoped that further algorithmic developments could support the practice of radiology among other specialities \citep{topol2019converge}. Deep learning is a sub-type of artificial intelligence that utilises multiple layers of analysis to process an image. A variety of applications of deep learning are postulated \citep{lieman127, wang2017cxr8, chilamkurthy2018deep}, and one study has shown this to be a useful approach in automated VS segmentation \citep{shapey2019artificial} in terms of both time and accuracy. Despite the accuracy of automated approaches, interactive corrections may continue to play a role even with deep learning due to the lack of adaptability of automated methods to the specific imaging sequences and protocols used clinically \citep{wang2019interact}. The next steps are to further analyse this methodology and work towards clinical translation. 

The findings of this study may also be applied more widely to semi-automated segmentation of other neuroimaging data. Some participants felt that manual segmentation could not be matched in terms of performance if plenty of time was spent. The participants did not have specific expertise in the diagnosis or management of VS, aside from the neurosurgeon. We would expect that similar results, in terms of qualitative findings, may be present in other applications; for instance tumour segmentation for glioma. We would recommend that semi-automated segmentation is used as a supportive measure to other standard approaches in neuroimaging segmentation. 

\section{Conclusion}
%\label{sec:conclusion}
Gains are being made in the machine learning and medical imaging fields. Machine learning applications are now performing comparably with their manual counterparts. However, a finding of this study was that even the state-of-the-art machine learning tools may not yet be fully ready for clinical roll out in segmentation of vestibular schwannoma.
Users found the tools to be fast and accurate, but at times unpredictable and frustrating to use. There were limitations in the study, including the small sample size in terms of participants, particularly those with experience in segmentation, and in the number of scans segmented. This makes conclusions difficult to draw. The strengths of this study lie in the joint use of both qualitative and quantitative methods, which were employed to address the clinical applicability of algorithms. Unpredictability of algorithm behaviour and lack of transparency with algorithmic methods are cited as being key issues. To remedy this, developers should focus on involving groups with a variety of backgrounds and expertise in the development process, to ensure clinical and research applicability. 

\begin{acknowledgements}
This work was supported by Wellcome [203145Z/16/Z, 203148/Z/16/Z, WT106882] and EPSRC [NS/A000050/1, NS/A000049/1] funding.
TV is supported by a Medtronic/Royal Academy of Engineering Research Chair [RCSRF1819\textbackslash7\textbackslash34].
%
% Authors must disclose all relationships or interests that 
% could have direct or potential influence or impart bias on 
% the work: 

\noindent\textbf{Conflict of interest}
An academic license was provided by ImFusion for the use of ImFusion Labels. Besides this, the authors have no conflicts of interest to declare.
%
% The authors declare that they have no conflict of interest.

\noindent\textbf{Human and Animal Rights}
There were no human or animal studies conducted in this work.

\noindent\textbf{Informed Consent}
There was no informed consent or IRB study required for the work reported in this manuscript.
\end{acknowledgements}

% BibTeX users please use one of
% References should be in square brackets
% https://www.springer.com/journal/11548/submission-guidelines?IFA#Instructions%20for%20Authors_References
%\bibliographystyle{spbasic}      % basic style, author-year citations
\bibliographystyle{spmpsci}      % mathematics and physical sciences
\bibliography{2020-SemiautoSeg-McGrath}   % name your BibTeX data base

\newpage\FloatBarrier
\appendix
\section*{Appendix}

\begin{table}[ht]
% table caption is above the table
\caption{NASA Task Load Index.
Hart and Staveland’s NASA Task Load Index (TLX) method assesses
work load on five 7-point scales. Increments of high, medium and low estimates for each point result in 21 gradations on the scales.}
\label{tab:TLX_questionnaire}       % Give a unique label
% For LaTeX tables use
\begin{tabular}{l}
\hline\noalign{\smallskip}
How mentally demanding was the task?   \\
\noalign{\smallskip}
%\hline\noalign{\smallskip}
How physically demanding was the task? \\
\noalign{\smallskip}
%\hline\noalign{\smallskip}
How hurried or rushed was the pace of the task? \\
\noalign{\smallskip}
%\hline\noalign{\smallskip}
How successful were you in accomplishing what
you were asked to do? \\
\noalign{\smallskip}
%\hline\noalign{\smallskip}
How hard did you have to work to accomplish
your level of performance? \\
\noalign{\smallskip}
%\hline\noalign{\smallskip}
How insecure, discouraged, irritated, stressed,
and annoyed were you? \\
\noalign{\smallskip}\hline
\end{tabular}
\end{table}

\begin{table}[ht]
% table caption is above the table
\caption{Interview questions for qualitative comparison of the two software libraries}
\label{tab:questionnaire}       % Give a unique label
% For LaTeX tables use
\begin{tabular}{l}
\hline\noalign{\smallskip}
Was the segmentation in each software to your satisfaction? \\
\noalign{\smallskip}
%\hline\noalign{\smallskip}
Overall, how did you find each software? \\
\noalign{\smallskip}
%\hline\noalign{\smallskip}
What would you add or remove from each software to improve them? \\
\noalign{\smallskip}
%\hline\noalign{\smallskip}
How did you find the study? \\
\noalign{\smallskip}\hline
\end{tabular}
\end{table}

\begin{table}[htbp!]
 %table caption is above the table
\caption{Mean segmentation accuracy values for each scan in ITK-SNAP and ImFusion}
\label{tab:seg_accuracy}       % Give a unique label
 %For LaTeX tables use
\begin{tabular}{lllllll}
\hline\noalign{\smallskip}
& \multicolumn{3}{l}{ITK-SNAP} & \multicolumn{3}{l}{ImFusion} \\
\hline\noalign{\smallskip}
Tumour Identifier & Dice & RVE & ASSD & Dice & RVE & ASSD  \\
\noalign{\smallskip}\hline\noalign{\smallskip}
VS\_1 & 0.882 & 0.094 & 0.457 & 0.885 & 0.114 & 0.424 \\
VS\_2 & 0.893 & 0.110 & 0.398 & 0.890 & 0.043 & 0.422 \\
VS\_3 & 0.929 & 0.115 & 0.441 & 0.945 & 0.085 & 0.357 \\
VS\_4 & 0.903 & 0.178 & 0.379 & 0.925 & 0.056 & 0.311 \\
\noalign{\smallskip}\hline
\end{tabular}
\end{table}

\begin{table}[!hbt]
% table caption is above the table
\caption{Interview answers grouped by theme}
\label{tab:questionaire-analysis}       % Give a unique label
% For LaTeX tables use
\begin{tabular}{p{2cm}p{1.2cm}p{5.3cm}p{2cm}}
\hline\noalign{\smallskip}
\textbf{Theme} & \textbf{Software} & \textbf{Quotes} & \textbf{Prevalence}  \\
\noalign{\smallskip}\hline\noalign{\smallskip}
Performance discrepancy across tumours 
    & ImFusion
    & ‘Very good for clear-cut, simple tumours… [those which were] highly contrast enhancing, homogeneous,  with well-defined boundaries and minimal adjacent blood vessels.’
    
    ‘Complex tumours threw up errors which required a complete restart.’
    & All five participants (100\%) \\
\noalign{\smallskip}\hline\noalign{\smallskip}
Compromise between thoroughness and timing 
    & ITK-SNAP 
    & ‘I am a perfectionist… if we weren’t timed it would take me much longer.’
    
    ‘I made lots of small mistakes… but it would have taken too long to correct.’
    
    ‘It was very fiddly.’
    & Four out of five (80\%)  \\
\noalign{\smallskip}\hline\noalign{\smallskip}
Unpredictable outcome after drawing labels 
    & ImFusion 
    & ‘a one-trick pony… if you make alterations to the initial segmentation you may worsen it.’
    
    ‘if we wanted perfection… we would have to go back again and again.’
    
    ‘I do not know if the changes I make will improve of worsen the segmentation.’
    & Three out of five (60\%)  \\
\noalign{\smallskip}\hline\noalign{\smallskip}
Speed of segmentation
    & ImFusion 
    & ‘Much faster so it would be great for my work.’
    
    ‘The algorithm works very quickly.’
    & Four out of five (80\%) \\
\noalign{\smallskip}\hline\noalign{\smallskip}
UI and tools
    & Both
    & ‘[Using ImFusion] was a much nicer experience… and a sleek UI.’ 
    
    ‘[ImFusion] is better for visualization.’ 
    
    ‘[In ImFusion] I would like to have a paintbrush tool which draws and erases exactly what I want it to… there is too much prediction required… scribbles I make should not affect the whole segmentation.’ 
    & - \\
\noalign{\smallskip}\hline\noalign{\smallskip}
Study design 
    & -
    & ‘It was helpful to have someone here to explain and provide feedback.’ 
    
    ‘Would have been good to define the goal more clearly… do we want a very accurate segmentation or a rough volume estimate.’
    
    ‘You could have gone through all the tools I might need during the training phase.’
    & - \\

\noalign{\smallskip}\hline
\end{tabular}
\end{table}

\end{document}